\documentclass[prb,twocolumn,superscriptaddress,showpacs,amsmath,amssymb]{revtex4-1}

\usepackage{graphicx}
\usepackage{latexsym}
\usepackage{amsmath}
\usepackage{amssymb}
\usepackage{amsfonts}
\usepackage{color}
\usepackage{bm}
\usepackage{verbatim}
\usepackage{pagecolor,lipsum}

\begin{document}

 \title{  Anomalous  current  in diffusive  ferromagnetic Josephson  junctions. }

\date{\today}

 \author{M.A.~Silaev}
 \affiliation{Department of
Physics and Nanoscience Center, University of Jyv\"askyl\"a, P.O.
Box 35 (YFL), FI-40014 University of Jyv\"askyl\"a, Finland}

\author{I.V.~Tokatly}
\affiliation{Nano-Bio Spectroscopy group, Departamento de F\'{i}sica de Materiales,
Universidad del Pa\'{i}s Vasco, Av. Tolosa 72, E-20018 San Sebasti\'{a}n, Spain}
\affiliation{IKERBASQUE, Basque Foundation for Science, E-48011 Bilbao, Spain}

\author{F.S.~Bergeret}
\affiliation{ Centro de F\'{i}sica de Materiales (CFM-MPC), Centro
Mixto CSIC-UPV/EHU, and Donostia International Physics Center (DIPC), Manuel de Lardizabal 5, E-20018 San
Sebasti\'{a}n, Spain}

 \begin{abstract}
 We demonstrate that in diffusive superconductor/ferromagnet/superconductor (S/F/S) junctions 
 a finite, {\it anomalous},  Josephson current can flow even at zero phase difference between the S electrodes. 
 The  conditions for the observation of this effect are non-coplanar magnetization distribution and a broken 
 magnetization inversion symmetry of the
 superconducting current.  The latter  symmetry is intrinsic for the widely used quasiclassical approximation 
 and prevent previous works, based on this approximation, from obtaining the Josephson  anomalous current.
 We show that this symmetry  can be removed by introducing
 spin-dependent boundary conditions for the quasiclassical equations at the 
 superconducting/ferromagnet interfaces in diffusive systems.  
 Using this recipe we considered generic multilayer magnetic systems
 and determine the ideal experimental conditions in order to maximize the  anomalous current. 
 \end{abstract}

 \maketitle

 In its minimal form the current-phase relation (CPR) characterizing the dc Josephson effect reads
$ I (\varphi)= I_{c}\sin\varphi$, where $\varphi$ is the phase difference between 
 superconducting  electrodes and $|I_c|$ is the critical current that is the maximum supercurrent 
 that can flow through the junction\cite{Josephson,GolubovRMP}.
 Ordinary Josephson junctions  are characterized by $I_c>0$ 
 yielding the zero phase difference  ground state $\varphi =0$.  
 In certain cases, however, $I_c<0$ and the ground state corresponds to 
 $\varphi =\pi$. Such $\pi$-junctions  can be realized   
 for example in superconductor/ferromagnet/superconductor (SFS) 
 structures \cite{BuzdinPi,Ryazanov2001,Ryazanov2006,BuzdinRMP}, 
 Josephson junctions with  non-equilibrium normal metal interlayer \cite{KlapwijkSNS}, 
 $d$-wave superconductors\cite{Dwave},  semiconductor nanowires \cite{NazarovNature2006}, 
 gated carbon nanotubes\cite{CarbonNanotube} or multi-terminal Josephson systems \cite{Vischi2016}. 
 $\pi$-junctions  has  being suggested   
 for building scalable superconducting digital and quantum logic \cite{Cryogenic_computers,RyazanovNatPhys2010,BirgeNatPhys}.
    
 As for $\varphi=0,\pi$ junctions, no physical argument speaks against 
 a  CPR of the form
 \cite{Buzdin}
 \begin{equation} \label{Eq:CPRgen}
 I (\varphi)= I_{c}\sin(\varphi+\varphi_0)\; .  
 \end{equation}
 with an arbitrary phase shift $\varphi_0\neq \pi n$ and Josephson energy 
 $E_J = -I_c \cos(\varphi+\varphi_0)$.  In such a case the ground state  corresponds to 
 $\varphi=-\varphi_0$ and  a finite supercurrent
  at zero phase difference  $I_{an} = I_{c}\sin\varphi_0$, termed the {\it anomalous current}.
 This effect, referred as the anomalous Josephson effect (AJE), 
  takes place only in systems with a broken time reversal symmetry. 
 
 The AJE has been predicted 
 in junctions which combine conventional superconductors with magnetism and spin-orbital interaction 
 \cite{Buzdin, Zazunov, Zazunov2, Reynoso, Nesterov, KonschellePRB2015, BergeretTokatly, Bobkova, NazarovTokoyama},  
 between unconventional superconductors \cite{AJEpairing},
 and topologically non-trivial superconducting leads\cite{Loss}. In the presence of magnetic flux piercing the normal interlayer the 
   superconducting proximity currents are generated which naturally leads to the phase shift of CPR \cite{Dolcini,AlidoustLinder}. 
 Experimentally, a  $\varphi_0$-junction has been reported in nano-wire quantum dot\cite{Szombati} 
 controlled by an external magnetic field and an electrostatic gate. 
  
Another type of systems predicted to exhibit  the AJE are  conventional 
SFS junctions with a non-homogeneous magnetization texture
\cite{NazarovBraude,ajeEschrig,BuzdinMironov,LiuChan3Layer,ajeMargaris,LinderKulagina,MoorVolkov1,MoorVolkov2}. 
In such systems the current is a functional of the magnetization  
distribution  ${\bm M}$,  $I=I ( \varphi, {\bm M} ) $. Time-inversion symmetry dictates that 
 $I ( \varphi, {\bm M} ) = - I ( -\varphi, -{\bm M} )$. If the system has 
 an additional magnetization inversion symmetry such that 
 \begin{equation} \label{Eq:AdditionalSymmetry}
 I( \varphi, {\bm M} ) = I( \varphi, -{\bm M} ),
 \end{equation}  
 then  $I ( \varphi, {\bm M} ) = -I(- \varphi,{\bm M} )$  and obviously the system does not exhibit 
 the AJE. In other words, it is necessary to break the symmetry (\ref{Eq:AdditionalSymmetry}) in order to 
obtain the $\varphi_0$ state. 

For example, for any coplanar magnetization distribution exists a global SU(2) 
 spin rotation such that flips the direction of ${\bm M}$, and the condition (\ref{Eq:AdditionalSymmetry}) is fulfilled.
For this reason, the AJE requires a non-coplanar magnetization texture. This explains the AJE predicted for 
 ballistic S/F/F/F/S systems with non-collinear magnetizations\cite{LiuChan3Layer,ajeMargaris,LinderKulagina}. 
 The anomalous current obtained in those works
  shows rapid oscillations as a function of the ferromagnetic thickness.  
  These oscillations result from the Fabry-Perot interference of electronic waves reflected at the S/F and 
  F/F interfaces.  
    
In diffusive SFS  structures, as those used in experiments \cite{Ryazanov2001,Ryazanov2006,RyazanovNatPhys2010,KontosPRL2002,Birge,Robinson}
 the impurity scattering randomizes directions of electron propagation
and hence one expects  the suppression of the rapidly oscillating anomalous current. 
Studies, based on quasiclassics, of the diffusive Josephson junctions through various non-coplanar
 structures including helix \cite{Helix}, magnetic vortex\cite{Kalenkov} and 
 skyrmion\cite{LinderSkyrmion} have shown no AJE.   
In contrast, in diffusive systems with  half-metallic elements \cite{NazarovBraude,BuzdinMironov} 
and in junctions between magnetic superconductors with spin filters\cite{MoorVolkov1,MoorVolkov2}
a finite anomalous current has been predicted.  From this apparent contradiction, the general 
condition for the AJE in diffusive systems still remains elusive.

In this letter  we show for the first time that the AJE is a robust effect that can exists in any diffusive SFS systems 
with non-coplanar magnetization textures under quite general conditions. 
We demonstrate that  the reason why anomalous currents have not been found in  previous studies on diffusive SFS systems is due to the additional 
 magnetization  inversion symmetry (\ref{Eq:AdditionalSymmetry}) that the quasiclassical approximation\cite{Eilenberger,Usadel} has 
with respect to the original Hamiltonian  and that prevents the description of the AJE in ferromagnetic junctions. 
In a second part of the letter  we consider a  spin-filter at the S/F interfaces 
and demonstrate the existence of  anomalous currents in diffusive SF structures. This allow us to study the AJE 
 without having to renounce the widely used quasiclassical approximation\cite{BuzdinRMP,BergeretRMP}.

 We start by analyzing the inherent symmetries of the Usadel equation, which is a  
  diffusion-like equation for the  
 quasiclassical Green functions (GF). 
 In the Matsubara representation it has the form \cite{Usadel,BuzdinRMP,BergeretRMP} 
 \begin{equation} \label{Eq:Usadel}
 D\nabla (\check g \nabla \check g)=
 [\check\Delta + \hat\tau_3( \omega + i {\bm{\hat\sigma}}\bm{h} ), \check g], 
 \end{equation}
 where $[a,b]=(ab-ba)/2$, $\omega$ is the Matsubara frequency, 
 $\bm{h} ({\bm r})$  is the 
 exchange field which is parallel to the local magnetization ${\bm M}({\bm r})$,  
  ${ \bm{\hat\sigma}}=(\hat\sigma_1,\hat\sigma_2,\hat\sigma_3)$
  is the vector of Pauli matrices in spin space $\hat \sigma_{1,2,3}$ and $\hat \tau_{1,2,3}$ are the Pauli 
 matrices in Nambu space.
 The gap  matrix is defined as $\check \Delta = \hat\tau_1 \Delta e^{-i\hat\tau_3\varphi}$,
 where $\Delta$ and $\varphi$ are the magnitude and phase of the order parameter. 
   The 4$\times$4 matrix GF in spin-Nambu space can be written in the form, which takes into account 
 the general particle-hole symmetry of Eq.(\ref{Eq:Usadel}) 
  \begin{equation}\label{Eq:GFNambu}
 \check g = 
 \left(
  \begin{array}{cc}
    \hat g & \hat f \\
    \bar{ \hat f} & - \bar{\hat g} \\
  \end{array}
\right)
 \end{equation} 
with $2\times2$ components $\hat g$ and $\hat f$ in the spin space and the time-reversed operation  defined as 
$\bar X = \hat\sigma_2 X^* \hat\sigma_2$.  Eq. (\ref{Eq:Usadel}) is complemented by the normalization condition $\check g^2=1$.   
 
  We introduce the following transformation 
 \begin{equation}\label{Eq:Transformation}
 \check g_{new} = \hat\sigma_2 \hat\tau_1 \check g^* \hat\tau_1 \hat\sigma_2 \; , 
 \end{equation}
 which is  a combination of two 
 transformations $g_{new}={\cal T}\Theta g \Theta^\dagger  {\cal T}^\dagger$: the
 time reversal  transformation, ${\cal T}= i\hat\sigma_2 {\cal K}$,  with ${\cal K}$ being the complex conjugate operation, 
 and the transposition of the electron and hole blocks of g,  $\Theta=\hat\tau_1$.
  Applying the transformation (\ref{Eq:Transformation}) to the Usadel Eq.(\ref{Eq:Usadel}) one obtains that 
 \begin{equation} \label{Eq:Symmetry}
 \check g_{new} (\omega, {\bm h})= \check g (-\omega, -{\bm h}) .
 \end{equation}
 On the other hand, the current is expressed as:
  \begin{equation} \label{Eq:Current}
 {\bm j} = i \frac{\sigma_n}{8e} \pi T \sum_{\omega=-\infty}^{\infty} {\rm Tr}\;\hat\tau_3 \check g \nabla \check g,
 \end{equation}
  where $\sigma_n=e^2N_F D$ is the normal metal conductivity and  $N_F$ is the density of states at the Fermi level. 
  The summation is done over Matsubara frequencies  $\omega=\pi T(2n+1)$, where $n$ is the integer number and $T$
  is the temperature. 
 It follows from Eqs. (\ref{Eq:Transformation}-\ref{Eq:Symmetry})    
 that the  current is invariant with respect to the 
magnetization inversion,  ${\bm j} (\bm h) = {\bm j} (-\bm h)$,  as anticipated in Eq. (\ref{Eq:AdditionalSymmetry}). 
By combining this extra symmetry with the general time-reversal symmetry,
${\bm j} (\varphi, {\bm h}) = -{\bm j} (-\varphi, - {\bm h})$ one obtains  that 
${\bm j} (\varphi) = -{\bm j} (-\varphi)$ and hence within, the quasiclassical approach, the AJE cannot take place for any spatial dependence of the exchange
 field ${\bm h}({\bm r})$. On the other hand, we know from previous works  that anomalous current may be generated at least in ballistic 
 SFS junctions with non-coplanar configuration of the magnetization \cite{LiuChan3Layer,ajeMargaris,LinderKulagina}. What is the origin of the apparent contradiction between the explicit ballistic calculations in those references and the magnetization 
 reversal symmetry of the Usadel equation? Is the absence of AJE  a specific feature of diffusive systems or is there a deeper reason for the above symmetry? 

To answer these questions let us first recall the Bogoliubov-De Gennes (BdG) Hamiltonian:
\[
H_{BdG}=\left(\begin{array}{cc}
\xi-{\bm{\hat\sigma}}\bm{h} & \Delta\\
\Delta^{*} & -\xi-{\bm{\hat\sigma}}\bm{h}
\end{array}\right)
\]
where $\xi=p^2/2m-E_F$ is the quasiparticle energy relative to the Fermi energy $E_F$.
 The general symmetries of the BdG  Hamiltonian are well known \cite{SchnyderRMP}. 
In the quasiclassical limit, which is equivalent to the Andreev approximation\cite{KonschellePRL}, transport properties are determined by particles living exactly at the Fermi surface. 
In the BdG Hamiltonian this corresponds to the $\xi=0$ case. In this, and only in this case, the BdG Hamiltonian 
acquires an additional symmetry with respect to the transposition of the electron and hole blocks, namely $\hat\tau_1H_{BdG}(\xi=0,\varphi,\bm h)\hat\tau_1= H_{BdG}(\xi=0,-\varphi,\bm h)$. According to Eq. (\ref{Eq:Transformation}) this symmetry together with the time-reversal operation leads to the invariance of the current under magnetization inversion. Obviously, this invariance is a general  feature of the quasiclassical theory, which holds true not only in the diffusive (Usadel) limit, but also for the full Eilenberger equation. In particular it explains why no AJE is obtained at the leading quasiclassical order in ballistic junctions with generic spin fields \cite{KonschellePRB2016}.
 
Clearly in real materials quantum effects always break this symmetry to a degree determined 
by the accuracy of quasiclassical approximation, which is the ratio $h/E_F$. 
Once this symmetry is broken the AJE may occur in any SFS system with arbitrary 
degree of non-magnetic disorder and non-coplanar magnetization distribution.  
The magnitude of the anomalous current will then be in leading order of the parameter $h/ E_F $. 
Typical experiments on SFS junctions showing the $\pi$-junction behavior,
used  weak ferromagnets \cite{Ryazanov2001,Ryazanov2006,KontosPRL2001}, for which  $h/ E_F \ll 1$. 
Therefore, at first glance, the AJE is hardly expected to be observed in these structures.

 This conclusion is however not fully correct, and there is indeed
 a way to enhance the anomalous Josephson currents 
 in systems with weak ferromagnets if one  introduces 
 spin-filtering tunnel barriers at the S/F interfaces,{\it i.e.} barriers with spin-dependent transmission for up and down spins.
 As we show below such barriers breaks the  quasiclassical symmetry, Eq. (\ref{Eq:AdditionalSymmetry}) 
and can lead to a measurable AJE in realistic SFS junctions. 
  
 Spin-filtering barriers are  described by the generalized  Kuprianov-Lukichev boundary conditions \cite{KL}, that include  spin-polarized 
 tunnelling at the SF interfaces \cite{BergeretVerso,EschrigLinderBC} 
 \begin{equation}
 \gamma \check g \partial_n \check g = [\check\Gamma \check g_{S} \check\Gamma^\dagger, \check g ].
 \label{Eq:KupLuk}
 \end{equation} 
  Here  $\partial_n = ({\bm n} \cdot\nabla )$ is the normal derivative at the surface,
  $\gamma = \sigma_n R$ is the parameter describing the barrier strength, $R$ is
 the normal state tunneling resistance 
 per unit area, and $\check g_S$  
 is the Green function of the superconducting electrode. We assume that the magnetization 
 of the  barriers points in ${\bm m}$ direction.  The spin-polarized tunneling matrix 
 has  the form $\check\Gamma= t\hat\sigma_0\hat\tau_0 + u(\bm { m \hat\sigma})\hat\tau_3 $, with 
 $t = \sqrt{( 1+ \sqrt{1-P^2} )/2} $,  $u = \sqrt{( 1 - \sqrt{1-P^2} )/2} $ and $P$ 
 being the spin-filter efficiency of the barrier that ranges from $0$ (no polarization)  to $1$ (100\% filtering efficiency).
  
By applying the transformation (\ref{Eq:Transformation}) to Eq. (\ref{Eq:KupLuk}) one can 
easily check the sign of the barrier polarization does not change and hence  
 \begin{equation}
 \label{Eq:QuasiclassicsSym}
 I (\varphi, \bm h, \bm P ) = I (\varphi, - \bm h , \bm P ) ,
 \end{equation}
 where $\bm P = P \bm m$. On the other hand,  the time-reversal transformation flips all the 
 magnetic moments including the exchange field and the barrier polarizations 
 \begin{equation} \label{Eq:TimeRevSymm1}
 I (\varphi, \bm h , \bm P ) = - I (-\varphi, - \bm h , - \bm P ).
 \end{equation}
 Combining Eqs. (\ref{Eq:QuasiclassicsSym},\ref{Eq:TimeRevSymm1}) we see, 
 that in principle, $ I (\varphi, \bm h, \bm P )\neq-I (-\varphi, \bm h, \bm P )$ 
 and  the zero-phase difference current at $\varphi =0 $ is not prohibited by symmetry. 
 
 \begin{figure}[h!]
 \centerline{$
 \begin{array}{c}
 \includegraphics[width=0.9\linewidth]{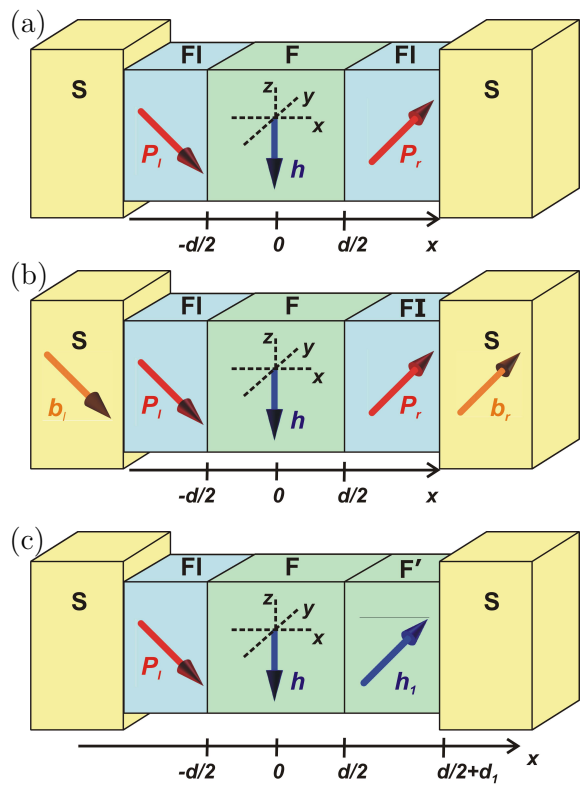} 
 \end{array}$}
 \caption{\label{Fig:SFSmodel} (Color online) 
 Generic non-coplanar tri-layer SFS systems: 
 (a) Non-collinear spin-filtering barriers (FI) with polarizations ${\bm P}_{r,l}$ 
 and metallic ferromagnetic layer (F) with exchange field ${\bm h}$. 
 (b) The same configuration as in (a) and Zeeman fields ${\bm b}_{r,l}$ in superconducting electrodes. 
 (c) Spin-filtering barrier with polarization ${\bm P}$ and two layers of metallic ferromagnet 	with non-collinear 
  magnetizations ${\bm h}_1$ (F) and ${\bm h}_2$ (F'). }
 \end{figure}

From this analysis is clear that the general features of the CPR can be deduced from the symmetry relations
(\ref{Eq:QuasiclassicsSym},\ref{Eq:TimeRevSymm1}).
First we consider the S/FI/F/FI/S structure of  Fig. \ref{Fig:SFSmodel}a.
 Here FI stands for the spin-filtering barriers with magnetizations ${\bm P}_{r,l}$
 and F is the mono-domain weak ferromagnet with exchange field ${\bm h}$.
  From previous works\cite{LiuChan3Layer,ajeMargaris,LinderKulagina} 
one  would expect that the anomalous current is proportional to the spin chirality $\chi =\bm{h}\cdot ({\bm P}_r\times{\bm P}_l) $.
However,  this term is prohibited by the symmetry (\ref{Eq:QuasiclassicsSym}) because of the change of sign of $\chi$. 
Instead,  one can construct the scalar  $I_{an}\propto ({\bm P_{r,l} \bm{ h}})\chi$ which
 is invariant to the sign change of $\bm{ h}\to -\bm{ h}$ and therefore is robust 
to the quasiclassical symmetry (\ref{Eq:QuasiclassicsSym}).  In such a case  the anomalous current is finite if  all vectors
are non-collinear  and in addition ${\bm h}$ has a component parallel to at least one of the magnetizations  ${\bm P}_{l,e}$.
 
To get an agreement with the results based on  the 
 Bogolubov - de Gennes calculations \cite{ajeMargaris}, 
 that yield $I_{an}\neq 0$ for any non-coplanar spin texture and has to take  
  into account the magnetic proximity effect
 \cite{Tokuyasu,BVE_inv,BLM_inv, Moodera} that induces an effective exchange  field ${\bm b}_{r}$ and ${\bm b}_{l}$ in the superconducting electrodes
 [Fig.(\ref{Fig:SFSmodel})b].
  In this case we define the chiralities   $\chi_{l,r} = {\bm P}_{l,r}\cdot ({\bm b}_{r,l}\times \bm{ h})$.
 which are invariant respect to the quasiclassical symmetry since they 
 contain two exchange fields changing signs under the transformation (\ref{Eq:QuasiclassicsSym}). 
 Thus, in this case the AJE is expected to be proportional to a linear combination of the chiralities $\chi_{l,r}$. 
 
   A similar behavior can be obtained for the structure  shown in  Fig.\ref{Fig:SFSmodel}c. 
 It is a S/FI/F/F$^\prime$/S junction with non-coplanar configuration of the one barrier polarization 
 ${\bm P}_l$ and two ferromagnetic layers $\bm h$ and ${\bm h}_1$. 
 In this case the chirality $(\bm P_l \times \bm {h}_1) \bm{h} \neq 0 $ is invariant 
 under the symmetry (\ref{Eq:QuasiclassicsSym}) thus allowing for the existence of the AJE.     
  
 To quantify  these  effects  we calculate the CPR analytically 
 focusing on the weak proximity effect in the F layer that allows for a linearization of 
 the Usadel equation with respect to the anomalous Green's function\cite{BergeretRMP}. 
 The latter can be written 
 as a superposition of the scalar singlet amplitude $f_s$ and
 the vector of triplet states ${\bm f}_t= (f_x,f_y,f_z)$, 
 $\hat{f}=f_s\hat\sigma_0+{\bm f}_{t} \bm {\hat \sigma}$. From Eq. (\ref{Eq:Usadel}) we get the following system of equations  for  $\omega >0$
 \begin{align}\label{Eq:UsadelSinglet}
 & (D\nabla^2-2\omega) f_s -i {\bm h} {\bm f}_t = 0\\ \label{Eq:UsadelTriplet}
 & (D\nabla^2-2\omega) {\bm f}_t - if_s {\bm h} =0 \; . 
 \end{align}
 supplemented by the linearized boundary condition obtained\cite{SM} from Eq. (\ref{Eq:KupLuk})
 in case the possible exchange field in the superconductor is parallel to the barrier polarization ${\bm b}\parallel {\bm P}$
 \begin{equation}\label{bc0}
 \gamma \partial_n \hat f = -[\hat G_s \bm {P\hat\sigma}, \hat f] - 
 \{ \hat G_s, \hat f \} + 
 \sqrt{1-P^2} \hat F_s ,
 \end{equation}  
 where $\{a,b\}=(ab+ba)/2$, $\hat G_s$ and $\hat F_s$ are the the normal and anomalous GF in the superconducting electrodes.
 The first term in the right hand side of Eq.(\ref{bc0}) is novel as compared to the boundary conditions for non-magnetic interfaces ($P=0$). 
 This term provides a $\pi/2$ phase rotation of the triplet superconducting components  non-collinear with 
 the barrier  polarization  ${\bm P}$. It is precisely this phase rotation that may lead to  an effective shift of the phase difference between 
 the Cooper pairs across the junction resulting in the AJE. 
 
 We calculate the amplitude $I_{an}$ 
 for the structures shown in Fig.(\ref{Fig:SFSmodel}) in the practically relevant regime when the coherence length in
 the middle ferromagnetic layer $\xi_F = \sqrt{D/h}$ is much shorter than that in a  normal metal 
 $\xi_N = \sqrt{D/T}$. The analytical result can be obtained by  assuming that the length 
 $d$ of the junction is $\xi_F \ll d\ll \xi_N$. Under such conditions the Josephson current is mediated 
 by long-range triplet superconducting correlations (LRTSC)\cite{BergeretRMP} since short-range modes  decay over $\xi_F$.  
  
 For the S/FI/F/FI/S structure shown in Fig. \ref{Fig:SFSmodel}a
  we neglect the magnetic proximity effect 
  and assume the  bulk GF in the S electrodes
   $\hat G_s = \omega/\sqrt{\omega^2 +|\Delta|^2} \equiv G_0 $,  
  $\hat F_s = \Delta G_0/\omega \equiv F_0 $.
     Then  
 the anomalous current is  \cite{SM}
 \begin{equation} \label{Eq:AnCurrent}
 \frac{eR I_{an}}{2\pi} = \chi ({\bm { h}} \bm {\bar P} ) \sqrt{(1-P_l^2)(1-P_r^2)} 
 \frac{\xi_F^2 T}{\gamma^5h^2d^2} \sum_{\omega>0} \frac{F_0^2G_0^2}{8k_\omega^4} 
 \end{equation}
 where $\bm {\bar P}= {\bm P}_r+ {\bm P}_l$ and 
 $k_\omega = \sqrt{\omega/D}$.  
 As expected for this case, the anomalous current is  
 proportional to $(\bm{ h}\bm {\bar P})\chi$, where 
 $\chi =\bm{ h}\cdot ({\bm P}_r\times{\bm P}_l) $ is the  spin chirality. 
 It is important to note that  
 the usual contribution to the Josephson current $I_0=I(\varphi=\pi/2)$  
 determined by the  LRTSC is proportional to
 $I_0\propto \gamma^{-4}$, and hence it dominates over the anomalous one, $I_0 \gg I_{an}\propto \gamma^{-5}$.
  
  If we now take the inverse proximity effect into account,  and  assume 
  effective exchange  fields ${\bm b}_{r}$ and ${\bm b}_{l}$ in the
  superconductors [Fig.(\ref{Fig:SFSmodel} )b], we obtain\cite{SM}
\begin{equation} \label{Eq:Anomalous2} 
  \frac{eR I_{an}}{2\pi} = (\chi_l-\chi_r) \sqrt{(1-P_r^2)(1-P_l^2)}   
 \frac{\xi_F}{\gamma^3 hd }\sum_{\omega>0}\frac{TF_0^\prime F_0 G_0 } { 2\sqrt{2} k_\omega^2} 
   \end{equation}
   where the chiralities $\chi_{r,l}$ are defined above and $F_0^\prime = dF_0/d\omega$.
  The usual current carried by the LRTSC is, to the lowest order in transparency, given by \cite{SM}
     $I_0\propto \gamma^{-2}( {\bm b}_{r\perp}{\bm b}_{l\perp})$, where 
  $  {\bm b}_{r,l\perp} = {\bm b}_{r,l} - {\bm h}({\bm b}_{r,l} {\bm { h}})/h^2$
  are the projections onto the plane perpendicular to the exchange field  ${\bm h}$. 
  In contrast to the previous example,  $I_0$ is given by the lower order in 
  $\gamma^{-1}$ since the LRTSC can tunnel directly from the 
  superconducting electrodes modified by the exchange fields ${\bm b}_{r,l}$. 
 Hence in general if $( {\bm b}_{r\perp}{\bm b}_{l\perp}) \neq 0$
 the anomalous current (\ref{Eq:Anomalous2}) 
 is a factor  $\xi_F/\gamma\ll 1$ smaller than $I_0$. 
 However if either  $b_r$ or $b_l$ vanishes, then  $I_0\propto \gamma^{-4}$ and the anomalous current dominates. This leads   
 to a large  AJE with  $I=I_{an}\cos\varphi$ so that the Josephson current has its  maximal value at zero phase difference. 
          
 In the practice the situation equivalent to e.g. $b_r=0$ and $b_l\neq 0$ 
  can be realized in a S/FI/F/F$^\prime$/S junction [Fig.\ref{Fig:SFSmodel}c]. 
  If  the middle F layer satisfies the above condition, $\xi_F \ll d \ll \xi_N$,  but the
   right layer, F$^\prime$ is   short enough such that  $d_1\ll \xi_F$, 
  the zero-phase difference current is given by\cite{SM} 
  \begin{equation}\label{Eq:CurrentFFIan1} 
 \frac{eR I_{an}}{2\pi} =  \chi \sqrt{1-P_l^2}  \frac{d_1 T}{4\gamma^3 h^2 d } 
 \sum_{\omega>0} \frac{F^2_0 G_0}{k_\omega^2}  
 \end{equation} 
where $\chi= (\bm P_l \times \bm { h}_1) \bm{ h} \neq 0 $.
 As in our first example, the usual component of the current
   is  proportional to \cite{SM} $I_0\propto \gamma^{-4}$ and therefore $I_{an}\gg I_0$ . 
 This type of S/FI/F/F$^\prime$/S structure provides the maximal AJE since 
 the anomalous current  is of the same order of the critical one  $I_{an}\sim I_c$. 
 
All  previous results are strictly speaking valid in the quasiclassical limit in which $h/E_F\ll1$.
 However, in the case of strong ferromagnets, $h/E_F\lesssim 1$, the difference between Fermi velocities for spin up and down electrons
can be described by an effective spin-filtering effect at the S/F interfaces, and therefore they also apply for for ballistic systems and strong ferromagnets.   
         
 To summarize,  the proposed mechanism for the  AJE and $\varphi_0$ ground states in SFS structures 
 is rather generic and exists in any system with  non-coplanar magnetization configuration.
 This conclusion is in contrast to a number of previous studies  which did not obtain anomalous currents in diffusive
 and ballistic systems in the framework  of quasiclassical approximation.
 We clarify this apparent controversy by demonstrating   that the absence of AJE within quasiclassics
 is due to an additional symmetry  which is only exact at the Fermi level.
 In order to restore the symmetries of the original Hamiltonian we have considered
  spin-filtering boundary conditions to the Usadel equations and found analytical expressions for the anomalous current 
   in different geometries. Our results show that in structures as those shown in Figs. \ref{Fig:SFSmodel}b,c, the amplitude of the anomalous current 
comparable to   critical one $I_{an}\sim I_c$, and therefore  the AJE may be observed  in such junctions. 

{\it Acknowledgements} We thank T. T Heikkil\"a for stimulating discussions.  
M.A.S. acknowledges discussions with A. Mel'nikov, I. Bobkova and A. Bobkov. 
The work of F.S.B.  is supported by Spanish Ministerio de Economia y Competitividad (MINECO) through Project No. FIS2014-55987-P.
 I.V.T. acknowledges support from the Spanish Grant No. FIS2016-79464-P
  and from the 'Grupos Consolidados UPV/EHU del Gobierno Vasco' (Grant No. IT578-13). The work of M.A.S. was supported by the Academy of Finland.

 \section{Supplementary Material: Derivation of current-phase relations.}

 Here we derive analytical expressions for the anomalous and usual Josephson current components in generic 
 trilayer SFS structures shown in Figs.(\ref{Fig:SFSmodel}).
 We use Usadel Eqs.(\ref{Eq:UsadelSinglet},\ref{Eq:UsadelTriplet}) with boundary conditions 
 obtained from the linearization of the Eq.(\ref{Eq:KupLuk}). 
 For the general spin structure of GF in the superconducting electrode
 the linearized  boundary condition can be written as follows
  \begin{equation} \label{Eq:bcLinGen}
 \gamma \partial_n\hat f=  \hat{{\cal F}_s} - (\hat{{\cal G}_s} \hat f + \hat f \bar{\hat{{\cal G}_s}})/2 
 \end{equation}
 where 
 \begin{align}
 \hat{{\cal G}_s} = t^2\hat G_s + u^2 (\bm {m \hat\sigma } \hat G_s  \bm {m \hat\sigma }) + 
 2ut \{\hat G_s, \bm {m \hat\sigma } \} \\
 \hat{{\cal F}_s} = t^2\hat F_s - u^2 (\bm {m \hat\sigma } \hat F_s  \bm {m \hat\sigma }) - 
 2ut [\hat G_s, \bm {m \hat\sigma } ] .
 \end{align}
  In the presence of exchange field ${\bm b}$ in the superconducting electrode
 the GF are 
   \begin{align}\label{Eq:Gs}
 \hat G_s = G_{0} - i (\bm{\hat \sigma} \bm b) dG_0/d\omega \\ \label{Eq:Fs}
 \hat F_s = F_{0} - i (\bm{\hat \sigma} \bm b) dF_0/d\omega .
 \end{align}   
   If the exchange field is collinear with the barrier polarization 
  ${\bm b}\parallel {\bm m}$
  the boundary condition (\ref{Eq:bcLinGen}) acquires the form of Eq.(\ref{bc0}).
    In the right hand side of Eq.(\ref{bc0}) the first and second terms are much smaller than the third one. 
 Both the first and second terms are 
 proportional to the small tunnelling parameter $\gamma^{-1}$ but have
 different symmetry. The third term can be safely neglected since it
 has the same symmetry as the left hand side and therefore does not 
 provide any qualitative corrections. 
 We keep the second term which is important to obtain anomalous Josephson effect.    
  
 To calculate  the charge current in the ferromagnetic layer we use the expression 
 \begin{equation} \label{Eq:ChargeCurrent}
 \bm j = \frac{2\sigma_n}{e} \pi T  \sum_{\omega>0} {\rm Im} ( f_s^* \nabla f_s - \bm f_t^*\nabla \bm f_t ) .
 \end{equation}
  which is obtained linarizing the general Eq.(\ref{Eq:Current}).
  
  \subsection{S/FI/F/FI/S structure}
 
   First of all we consider the simplest possible tri-layer non-coplanar structure 
  S/FI/F/FI/S, where FI stands for the spin-filtering barriers with magnetizations ${\bm P}_{r,l}$
  and F is the mono-domain weak ferromagnet with exchange field ${\bm h}$.
  We calculate the CPR for the structure shown in Fig.(\ref{Fig:SFSmodel})a
 assuming without loss of generality that the exchange  filed is ${\bm h} =h {\bm z}$ and 
 ${\bm P}_{r,l}$
 can have arbitrary directions. Then we have the Usadel equations in components: 
 \begin{align}\label{Eq:UsadelSinglet1}
 & D\nabla^2 f_s = i h f_z \\ \label{Eq:UsadelTriplet1}
 & D\nabla^2 {\bm f}_z = ihf_s  , 
 \end{align}
 \begin{align}\label{Eq:UsadelFx}
 & D\nabla^2 f_x = 2\omega f_x \\ \label{Eq:UsadelFy}
 & D\nabla^2 f_y = 2\omega f_y  , 
 \end{align}
 In Eqs.(\ref{Eq:UsadelSinglet1},\ref{Eq:UsadelTriplet1}) we neglected $\omega$ which is small compared to the exchange energy. 

 The boundary conditions at the left electrode $x=- d/2$ 
  \begin{align} 
  \label{Eq:fsLeft}
  &  \gamma \partial_x f_s = -F_0 \sqrt{1-P_l^2} e^{-i\varphi/2} \\
   \label{Eq:fzLeft}
   & \gamma \partial_x f_z = i G_0 (P_x^lf_y - P_y^lf_x) \\
    \label{Eq:fxLeft}
  &\gamma \partial_x f_x = i G_0 (P_y^lf_z - P_z^lf_y) \\
   \label{Eq:fyLeft}
  & \gamma \partial_x f_y = i G_0 (P_z^lf_x - P_x^lf_z)  
  \end{align}
  and at the right electrode $x=d/2$  
  \begin{align}
  \label{Eq:fsRight}
  & \gamma \partial_x f_s = F_0 \sqrt{1-P_r^2} e^{i\varphi/2} \\
  \label{Eq:fzRight}
  & \gamma \partial_x f_z = - i G_0 (P_x^rf_y - P_y^rf_x) \\
  \label{Eq:fxRight}
  &\gamma \partial_x f_x = - i G_0 (P_y^rf_z - P_z^rf_y) \\
  \label{Eq:fyRight}
  & \gamma \partial_x f_y = - i G_0 (P_z^rf_x - P_x^rf_z)  
  \end{align}
  
 Using the above boundary conditions and the general expression for current (\ref{Eq:ChargeCurrent})
 we get that 
 \begin{equation} \label{Eq:ChargeCurrentBoundary}
 \frac{eR I}{2\pi} = 
  \frac{\sqrt{1-P_r^2}}{\gamma} T\sum_{\omega>0}  F_0{\rm Im} [ f_s^*(d/2)  e^{i\varphi/2} ] .
 \end{equation}

 To simplify the derivation we assume that the length is $\xi_F\ll d\ll \xi_\omega $
 where $\xi_F = \sqrt{D/h}$ and $\xi_\omega = \sqrt{D/\omega}$ are the coherence lengths in 
 normal and ferromagnetic regions.
 
 {\it The to the first order in tunnelling $\gamma^{-1}$ } 
 we can calculate $f_s$ and $f_z$ near each interface independently
 without overlapping.   For example at $x=d/2$ we have 
 \begin{align} 
 \label{Eq:AnsatzFs}
 & f_s^{(1)} = A_{1+} e^{k_1(x-d/2)} +  A_{2+} e^{k_2(x-d/2)} \\
 \label{Eq:AnsatzFz}
 & f_z^{(1)} = A_{1+} e^{k_1(x-d/2)}-  A_{2+} e^{k_2(x-d/2)}
 \end{align}
 where $k_{1,2}^2 = \pm i h/D$.  Then we get 
 \begin{align}\label{Eq:Feqd}
 & f_s^{(1)}(d/2) = \sqrt{1-P_r^2} ( \xi_F/ \sqrt{2}\gamma ) F_0 e^{i\varphi/2} \\ \label{Eq:fZd}
 & f_z^{(1)}(d/2) = -\sqrt{1-P_r^2} ( \xi_F/ \sqrt{2}\gamma )  F_0 e^{i\varphi/2 + i\pi/2} 
 \end{align} 
 and 
 \begin{align}\label{Eq:Feqmd}
 & f_s^{(1)}(-d/2) = \sqrt{1-P_l^2} ( \xi_F/ \sqrt{2}\gamma ) F_0 e^{-i\varphi/2} \\ \label{Eq:fZmd}
 & f_z^{(1)}(-d/2) = -\sqrt{1-P_l^2} ( \xi_F/ \sqrt{2}\gamma )  F_0 e^{-i\varphi/2 + i\pi/2} .
 \end{align} 

 {\it To the next order in $\gamma^{-1}$ }
 we find corrections to $f_s$ using the boundary conditions (\ref{Eq:fsRight},\ref{Eq:fzRight}).
 The amplitudes $f_{x,y}$ change negligibly small and therefore can be calculated integrating the 
 Eqs.(\ref{Eq:UsadelFx},\ref{Eq:UsadelFy}) and using the boundary conditions (\ref{Eq:fxRight},\ref{Eq:fxLeft}): 
 \begin{align}
 & f_x - i \beta \bar{P}_z f_y = -i\beta \overline{P_y f_z}  \\
 & f_y + i \beta \bar{P}_z f_x =  i\beta \overline{P_x f_z}
 \end{align}  
 where
 \begin{eqnarray} \label{Eq:beta}
 & \beta & = G_0 \xi_\omega^2/(2\gamma d) \\
 & \overline{P_y f_z} & = P^r_y f_z(d/2) + P^l_y f_z(-d/2) \\ 
 & \bar{P}_z & =  P^l_z +  P^r_z .
 \end{eqnarray}
 Hence we obtain
 \begin{align} \label{Eq:bcFx}
 & f_x =- i \beta \overline{P_y f_z} - \beta^2 \bar{P_z} \overline{P_x f_z}  \\
 \label{Eq:bcFy}
 & f_y =  i \beta \overline{P_x f_z} - \beta^2 \bar{P_z} \overline{P_y f_z}
 \end{align}  
 For the anomalous current we need the second terms in Eqs.(\ref{Eq:bcFx},\ref{Eq:bcFy}) 
 so that
 \begin{align} \label{Eq:Aux}
 & P_x^r f_y - P_y^r f_x =  \\ \nonumber
 & \beta^2 \bar{P}_z ( P_y^r P_x^l - P_x^r P_y^l ) f^{(0)}_z(-d/2) + 
 i(other terms) .
 \end{align}
  
 Now we can insert Eq.(\ref{Eq:Aux}) to the boundary conditions (\ref{Eq:fzRight},\ref{Eq:fsRight}) 
 to find the 
 corrections to the component $f_s(d/2)$ needed to calculate the current (\ref{Eq:ChargeCurrentBoundary}). 
 We search the correction $\tilde{f}_s$, $\tilde{f}_z$ in the form (\ref{Eq:AnsatzFs},\ref{Eq:AnsatzFz})
 \begin{align}
 \label{Eq:fsRightCorr}
 & \gamma \partial_x \tilde{f}_s = 0 \\
 \label{Eq:fzRightCorr}
 & \gamma \partial_x \tilde{f}_z = -i G_0 \beta^2 \bar{P}_z ( P_y^r P_x^l - P_x^r P_y^l ) 
 f^{(1)}_z(-d/2)
 \end{align}
 which yields 
 \begin{align}
 & \tilde{A}_{2+}= - \tilde{A}_{1+} k_1/k_2 \\   \label{Eq:A1}
 & \tilde{A}_{1+} = -i G_0 \frac{\beta^2}{2\gamma k_1} \bar{P}_z ( P_y^r P_x^l - P_x^r P_y^l ) 
 f^{(1)}_z(-d/2).
 \end{align}  
 Therefore
 $\tilde{f}_s(d/2) = (1-k_1/k_2) \tilde{A}_{1+} = (1-i) \tilde{A}_{1+} $. Substituting
 Eq.(\ref{Eq:A1}) we obtain 
 \begin{align} \nonumber
 & \tilde{f}_s(d/2)  = \\
 & -(1+i) \frac{ G_0 \beta^2}{2\gamma k_1}\bar{P}_z ( P_y^r P_x^l - P_x^r P_y^l ) f^{(1)}_z(-d/2) = \\
 \nonumber
 &- G_0 \frac{\beta^2 \xi_F}{\sqrt{2}\gamma} \bar{P}_z ( P_y^r P_x^l - P_x^r P_y^l ) f^{(1)}_z(-d/2) 
 \end{align}
 where we used the relation 
 $$
 k^{-1}_1 = e^{-i\pi/4}\xi_F = \frac{(1-i)}{\sqrt{2}} \xi_F
 $$
  Using Eq.(\ref{Eq:fZmd}) we obtain 
  \begin{align} 
 & \tilde{f}_s(d/2) = i \bar{P}_z ( P_y^r P_x^l - P_x^r P_y^l ) \sqrt{1-P_l^2}\times \\ \nonumber
 & \frac{\beta^2 \xi_F^2}{2 \gamma^2} 
  G_0F_0 e^{-i\varphi/2} .
 \end{align}
 
  Finally, substituting this expression to the Eq.(\ref{Eq:ChargeCurrentBoundary}) for the current 
  we obtain the anomalous current amplitude
 \begin{align} \label{Eq:AnCurrentSupplementary}
 \frac{eR I_{an}}{2\pi} = \bar{P}_z ( P_x^r P_y^l - P_y^r P_x^l ) \sqrt{1-P_l^2} 
 \sqrt{1-P_r^2} \times \\  \nonumber
 \frac{\xi_F^2 T}{8d^2\gamma^5} \sum_{\omega>0} \frac{F_0^2G_0^2}{k_\omega^4} 
 \end{align}
  We can write the amplitude of the current (\ref{Eq:AnCurrentSupplementary}) in the coordinate-independent form 
  $$
  h^2\bar{P}_z ( P_x^r P_y^l - P_y^r P_x^l ) = ({\bm { h}} \bm {\bar P}) \chi 
  $$ 
  where $\chi = {\bm{ h}} ({\bm P}_r\times {\bm P_l})$ and $\bm {\bar P}= {\bm P}_r+ {\bm P}_l$
  \begin{align} \label{Eq:AnCurrent1}
 \frac{eR I_{an}}{2\pi} = \chi ({\bm { h}} \bm {\bar P} ) \sqrt{1-P_l^2} 
 \sqrt{1-P_r^2} \times \\  \nonumber
 \frac{\xi_F^2 T}{8\gamma^5 h^2d^2} \sum_{\omega>0} \frac{F_0^2G_0^2}{k_\omega^4} .
 \end{align}
 
 \subsection{S/FI/F/FI/S structure with exchange field in superconducting electrodes}  
 \label{Sec:SFIFFISexch}  
 
 Next let us consider the same S/FI/F/FI/S trilayer system but take into account the induced exchange field
 in superconducting electrodes ${\bm b}_{r,l}$ shown in Fig.(\ref{Fig:SFSmodel})b. 
 In this case one can compose the chirality as follows 
 $\chi_l = {\bm P}_l\cdot ({\bm b}_r\times {\bm h})$ or $\chi_r = {\bm P}_r\cdot ({\bm b}_l\times {\bm h})$ 
 which are both robust against the quasiclassical symmetry 
 since both $\bm h$ and $\bm b_{r,l}$ change sign. 
  
 In the presence of  effective exchange  fields ${\bm b}_{r}$ and ${\bm b}_{l}$ 
  GF in the superconducting electrodes are given by Eqs.(\ref{Eq:Gs},\ref{Eq:Fs})
  with ${\bm b}= {\bm b}_{r} ( {\bm b}_{l})$ for right (left) electrodes.
 Substituting these expressions into boundary conditions
  Eq.(\ref{bc0}) we obtain at the left electrode $x=- d/2$ 
    \begin{align} 
   \label{Eq:fsLeft1}
   & \gamma \partial_x f_s  = -F_0 \sqrt{1-P_l^2} e^{-i\varphi/2} \\
   \label{Eq:fzLeft1}
   & \gamma \partial_x f_z = \\ \nonumber
   &  i G_0 (P_{lx} f_y - P_{ly} f_x)
    + i\sqrt{1-P_l^2} b_{lz} F_0^\prime e^{-i\varphi/2}  \\
   \label{Eq:fxLeft1}
   &\gamma \partial_x f_x = \\ \nonumber
   &   i G_0 (P_{ly} f_z - P_{lz} f_y)  
    + i\sqrt{1-P_l^2} b_{lx} F_0^\prime e^{-i\varphi/2}  \\
   \label{Eq:fyLeft1}
   & \gamma \partial_x f_y = \\ \nonumber
   & - i G_0 ( P_{lx} f_z - P_{lz} f_x ) 
    + i \sqrt{1-P_l^2} b_{ly} F_0^\prime e^{-i\varphi/2}   
   \end{align}
    and at the right electrode $x=d/2$  
  \begin{align}
  \label{Eq:fsRight1}
  & \gamma \partial_x f_s = F_0 \sqrt{1-P_r^2} e^{i\varphi/2} \\
  \label{Eq:fzRight1}
  & \gamma \partial_x f_z = \\  \nonumber
  & - i G_0 (P_{rx} f_y - P_{ry} f_x) 
   - i \sqrt{1-P_r^2} b_{rz}F_0^\prime e^{i\varphi/2} \\
  \label{Eq:fxRight1}
  &\gamma \partial_x f_x = \\ \nonumber
  & - i G_0 (P_{ry} f_z - P_{rz} f_y)
    - i\sqrt{1-P_r^2} b_{rx}F_0^\prime e^{i\varphi/2} \\
  \label{Eq:fyRight1}
  & \gamma \partial_x f_y = \\ \nonumber
  &  i G_0 ( P_{rx} f_z-P_{rz} f_x )
   - i\sqrt{1-P_r^2}b_{ry} F_0^\prime e^{i\varphi/2}  
  \end{align}
 
 Using this boundary conditions the current is given by 
 \begin{align} \label{Eq:ChargeCurrentBoundary2} 
 & \frac{eR I}{2\pi} = \frac{\sqrt{1-P_r^2}}{\gamma} \times \\ \nonumber
 & T\sum_{\omega>0} 
 {\rm Im} \{ e^{i\varphi/2} [ F_0 f_s^* + i F_0^\prime ({\bm b}_r {\bm f}_t^*)  ]\} .
 \end{align}

  To calculate the anomalous Josephson current  we assume again the regime 
 when $  \xi_F \ll d \ll \xi_N $. 
 In this case we can substitute the long-range components $f_{x,y}$ by their averages given by 
 \begin{equation} \label{Eq:Average}
  (2d k_\omega^{2}) \bar f_i =  \partial_x f_i (d/2) - \partial_x f_i (-d/2) .
 \end{equation}
  Substituting the boundary conditions (\ref{Eq:fxLeft1},\ref{Eq:fyLeft1},\ref{Eq:fxRight1},\ref{Eq:fyRight1})
  to the Eq.(\ref{Eq:Average}) and neglecting the terms of the order $\gamma^{-3}$ we obtain 
 \begin{align} \nonumber                                      
 & i{\bm b}_r{\bm f}^*_t = \frac{G_0}{2d\gamma k_\omega^2} ( P_{rx} b_{ry} - P_{ry} b_{rx}) 
 f_z^*(d/2) +  \\ \nonumber
 & \frac{G_0}{d\gamma k_\omega^2} ( P_{lx} b_{ry} - P_{ly} b_{rx}) f_z^*(-d/2) - \\ \nonumber
 & \frac{F^\prime_0}{2d\gamma k_\omega^2} \sqrt{1-P_l^2}( b_{rx} b_{lx} + b_{ry} b_{ly} ) 
 e^{i\varphi/2} - \\ \nonumber
 & \frac{F^\prime_0}{2d\gamma k_\omega^2} \sqrt{1-P_r^2}( b_{rx}^2 + b_{ry}^2 ) e^{-i\varphi/2}
 \end{align}
 
 Thus the second term in the current Eq.(\ref{Eq:ChargeCurrentBoundary2}) is given by
 \begin{align} \label{Eq:res1}
 & {\rm Im} ( i {\bm b}_r{\bm f}_t^* e^{i\varphi/2} ) = \\
 & -\sqrt{1-P_l^2} \frac{F_0^\prime}{d\gamma k_\omega^2} ( b_{rx} b_{lx} + b_{ry} b_{ly} ) 
 \sin\varphi + \\ \nonumber
 & \sqrt{1-P_l^2} \frac{F_0G_0\xi_F}{2\sqrt{2}d\gamma^2 k_\omega^2} 
 ( P_{lx} b_{ry} - P_{ly} b_{rx} ) \cos\varphi
 \end{align}
   
  To find the contribution of the first term in the current Eq.(\ref{Eq:ChargeCurrentBoundary2}) 
  we need to calculate the generation of singlet component at the boundary $x=d/2$ by the 
  long-range triplet ones $f_{x,y}$. To find this we take into account only the first (red) term 
  in the l.h.s. of the boundary conditions (\ref{Eq:fzRight1})
  \begin{align}
  & \partial_x \tilde f_s = 0 \\
  & \gamma \partial_x \tilde f_z = - i G_0 ( P_{rx} \bar f_y - P_{ry} \bar f_x )
  \end{align}
  Using the general solution (\ref{Eq:AnsatzFs}, \ref{Eq:AnsatzFz} ) 
  we obtain 
  \begin{equation}
  f_s(0) = -\frac{\xi_F G_0}{\sqrt{2}\gamma} (P_{rx} \bar f_y - P_{ry} \bar f_x)
  \end{equation}
  Therefore we get 
  \begin{align} \label{Eq:res2}
  & {\rm Im} (e^{i\varphi/2} f_s^*) = \\ \nonumber 
  & - \sqrt{1-P_l^2} \frac{F_0^\prime G_0 \xi_F}
  {2\sqrt{2}d\gamma^2 k_\omega^2} ( P_{rx} b_{ly} - P_{ry} b_{lx} ) \cos\varphi
  \end{align}
  
  The anomalous current is given by Eq.(\ref{Eq:res2}) and second term in (\ref{Eq:res1})
 \begin{align} \label{Eq:Anomalous2Supplementary} 
 & \frac{eR I_{an}}{2\pi} = \\ \nonumber
 & (\chi_l-\chi_r) \frac{\sqrt{1-P_r^2}\sqrt{1-P_l^2}\xi_F}{2\sqrt{2}h d\gamma^3 }
 \sum_{\omega>0}\frac{TF_0^\prime F_0 G_0 } {k_\omega^2} ,
 \end{align}
 where the chiralities are given by $\chi_l = {\bm P}_l\cdot ({\bm b}_r\times \bm{ h})$ and  
 $ \chi_r = {\bm P}_r\cdot ({\bm b}_l\times \bm{ h})$.
  The usual current is given by
  \begin{align} \label{Eq:Usual2} 
 &\frac{eR I_0}{2\pi} = -( {\bm b}_{r\perp}{\bm b}_{l\perp}) \sqrt{(1-P_l^2)(1-P_r^2)}  \times \\
  &\frac{1}{ 2\gamma^2 d}\sum_{\omega>0} \frac{TF_0^{\prime 2}}{k_N^2} \; ,
 \end{align}
 It is is proportional to the product 
 of the components $  {\bm b}_{r,l\perp} = {\bm b}_{r,l} - ({\bm b}_{r,l} {\bm { h}})/h $ perpendicular to the exchange field in 
 the ferromagnetic interlayer ${\bm h}$. 
  
 In general if $b_r$ and $b_l$ are non-zero $I_{an} \ll I_0$. 
 However if either $b_r=0$ or $b_l=0$ the usual component of Josephson current is absent $I_0=0$.
 In this case we obtain the giant anomalous Josephson effect when the CPR is given by $I=I_{an}\cos\varphi$
 and the current is maximal at zero phase difference. 
 
 Physically the situation equivalent to the case when the exchange field in one of the superconducting 
 electrodes is absent can be realized in the setup with non-homogeneous non-collinear magnetization in the metallic layer shown in Fig.(\ref{Fig:SFSmodel})c.   
   
 \subsection{S/FI/F/F/S structure with non-collinear exchange field}
 
 We consider the  non-coplanar tri-layered structure shown in Fig.(\ref{Fig:SFSmodel})c consisting of spin filter and two metallic ferromagnets. 
  The boundary conditions at the left electrode $x=- d/2$ 
 \begin{align} 
 \label{Eq:fsLeftFF}
 & \gamma \partial_x f_s = - F_0 \sqrt{1-P_l^2} e^{-i\varphi/2} \\
 \label{Eq:fzLeftFF}
 & \gamma \partial_x f_z = i G_0 (P_x^lf_y - P_y^lf_x) \\
 \label{Eq:fxLeftFF}
 &\gamma \partial_x f_x = i G_0 (P_y^lf_z - P_z^lf_y) \\
 \label{Eq:fyLeftFF}
 & \gamma \partial_x f_y = i G_0 (P_z^lf_x - P_x^lf_z)  
 \end{align}
 and at the right electrode $x=d/2+d_1$  
 \begin{align}
 \label{Eq:fsRightFF}
 & \gamma \partial_x f_s = F_0 e^{i\varphi/2} \\
 \label{Eq:fzRightFF}
 & \gamma \partial_x \bm f_t = 0 . 
 \end{align} 
 
 To obtain the boundary conditions at $x=d/2$  
 we can integrate through the layer  $d/2<x<d/2+d_1$
 to obtain the effective boundary conditions at $x=d/2$ which read as 
 \begin{align}
 \label{Eq:fsRightFF1}
 & \gamma \partial_x f_s = F_0 e^{i\varphi/2} - i\frac{\gamma d_1}{ D}(\bm h_1\bm f_t) \\
 \label{Eq:fzRightFF1}
 & \gamma \partial_x {\bm f}_t = - i\frac{\gamma d_1}{D} {\bm h_1} f_s 
 \end{align} 
 and at $x=-d/2$
 \begin{align} 
 \label{Eq:fsLeft5}
 & \gamma \partial_x f_s  = - F_0 \sqrt{1-P_l^2} e^{-i\varphi/2} \\
 \label{Eq:fzLeft5}
 & \gamma \partial_x f_z  =  0 \\
 \label{Eq:fxLeft5}
 &\gamma \partial_x f_x  =  i G_0 (P_y^l f_z - P_z^l f_y) + G_0 f_x  \\
 \label{Eq:fyLeft5}
 & \gamma \partial_x f_y = i G_0 (P_z^l f_x - P_x^l f_z) + G_0 f_y
 \end{align}
  These boundary conditions are qualitatively similar to (\ref{Eq:fxRight1},\ref{Eq:fyRight1}, \ref{Eq:fzRight1}). 
              
 Boundary conditions (\ref{Eq:fsRightFF}) yield the current given by 
 \begin{align}\label{Eq:CurrentFF} 
 \frac{eR I}{2\pi} =  T\sum_{\omega>0} \frac{F_0}{\gamma} {\rm Im} 
 [ e^{i\varphi/2} f_s^* ] 
 \end{align} 
  where $f_s^*=f_s^*(d/2)$.
  To find the current we need to determine corrections $f_s$ with the help of 
  boundary  conditions (\ref{Eq:fsRightFF1})
  due to the triplet components generated at the $x=-d/2$ boundary. 
  In this way we search the corrections to the short-range solution 
  $\tilde f_s, \tilde f_z$ in the form (\ref{Eq:AnsatzFs}, \ref{Eq:AnsatzFz})   
  with the amplitudes determined by the boundary condition 
  (\ref{Eq:fsRightFF1},\ref{Eq:fzRightFF1}).
   Thus we obtain 
   \begin{equation} \label{Eq:fsFF}
   \tilde f_s = -i \frac{d_1\xi_F}{\sqrt{2}D} ({\bm h_1}\bm f_t)
   \end{equation}
   The components ${\bm f}_t$ to be substituted in Eq.(\ref{Eq:fsFF}) 
   can be found using the equation (\ref{Eq:Average})
 \begin{align} \nonumber
 & i{\bm h_1 }{\bm{ f}}^*_t =  -(h_{1x}^2 + h_{1y}^2) \frac{d_1}{D d k^2_\omega} f^*_s(d/2) 
 \\ \nonumber
 & +\beta (h_{1y}P_{x}^l - h_{1x}P_{y}^l) f^*_z(-d/2) 
 \\ \nonumber
 &- i\beta^2 P_{z}^l (h_{1x}P_{x}^l + h_{1y}P_{y}^l) f^*_z(-d/2) ,
 \end{align}   
 where $\beta$ is given by (\ref{Eq:beta}).
  Using Eqs.(\ref{Eq:Feqd},\ref{Eq:fZmd}) we obtain
 $$
 {\rm Im} [ e^{i\varphi/2} f_s^* ] = \frac{d_1\xi_F}{\sqrt{2}D}
 {\rm Im} [ ie^{i\varphi/2} ({\bm h_1}\bm f_t^*)]  .
 $$
 Thus the anomalous and usual parts of the current (\ref{Eq:CurrentFF}) are given by 
 \begin{align}\label{Eq:CurrentFFIan} 
 \frac{eR I_{an}}{2\pi} =  \sqrt{1-P_l^2} ( h_{1y}P_{x}^l - h_{1x}P_{y}^l) \frac{d_1 T}{2\gamma^2h}  
 \sum_{\omega>0} \beta F_0^2 \\ \label{Eq:CurrentFFI0} 
 \frac{eR I_{0}}{2\pi} =  \sqrt{1-P_l^2} ( h_{1x}P_{x}^l + h_{1y}P_{y}^l) \frac{d_1 T}{2\gamma^2h}  
 \sum_{\omega>0} \beta^2 F_0^2
 \end{align} 
 
 These expressions can be rewritten in the coordinate-independent form 
 \begin{align}\label{Eq:CurrentFFIan1Supplementary} 
 & \frac{eR I_{an}}{2\pi} =  \chi \sqrt{1-P_l^2}  \frac{d_1 T}{4\gamma^3 d } 
 \sum_{\omega>0} \frac{F^2_0 G_0}{k_\omega^2}  
 \\ \label{Eq:CurrentFFI01} 
 & \frac{eR I_{0}}{2\pi} =  \sqrt{1-P_l^2} ( \bm P_l\bm{ h}_{1\perp} ) \frac{d_1 T}{8\gamma^4d^2h}  
 \sum_{\omega>0} \frac{F^2_0 G^2_0}{k_\omega^4}  
 \end{align} 
 where the chirality is given by $\chi =  (\bm P_l \times \bm { h}_1) \bm{ h}$ 
 and $\bm { h}_{1\perp} = \bm { h}_{1} - \bm { h} (\bm { h}\bm { h}_{1})/h^2 $
 is the perpendicular component of the exchange field ${\bm h}_1$. 
 The usual current is given by the higher order corrections in the tunnel barrier 
 transparency $I_0\propto \gamma^{-4}$ than the anomalous one $I_{an}\propto \gamma^{-3}$.
 Therefore in the tunnelling limit $I_{an}\gg I_0$. 


\end{document}